\documentclass{article}
\usepackage[utf8]{inputenc}
\usepackage{hyperref}
\usepackage{amsmath}
\usepackage{blkarray}
\usepackage{multicol}
\usepackage{multirow}
\usepackage{array}
\usepackage{floatrow}
\usepackage{float}
\usepackage{subcaption}
\usepackage{amsthm}
\usepackage{caption}
\usepackage{amsthm}
\usepackage{comment}
\usepackage{graphics}
\usepackage{graphicx}
\usepackage{grffile}
\usepackage[backend=bibtex]{biblatex}
\title{Using sequential drift detection to test the API economy}

\author{Samuel Ackerman
\thanks{IBM Research, Haifa; samuel.ackerman@ibm.com} 
\and Parijat Dube \thanks{IBM Research, Yorktown Heights}
\and Eitan Farchi \thanks{IBM Research, Haifa}}

\date{\today}

\begin{document}

\newcommand{\Prob}{\textrm{Pr}}
\newcommand{\nth}[1]{#1^{\text{th}}}

\maketitle

\begin{abstract}
The API economy refers to the widespread integration of API (advanced programming interface) microservices, where software applications can communicate with each other, as a crucial element in business models and functions.  The number of possible ways in which such a system could be used is huge.  It is thus desirable to monitor the usage patterns and identify when the system is used in a way that was never used before.  This provides a warning to the system analysts and they can ensure uninterrupted operation of the system.  

In this work we analyze both histograms and call graph of API usage to determine if the usage patterns of the system has shifted.  We compare the application of nonparametric statistical and Bayesian sequential analysis to the problem.  This is done in a way that overcomes the issue of repeated statistical tests and insures statistical significance of the alerts.  The technique was simulated and tested and proven effective in detecting the drift in various scenarios.  We also mention modifications to the technique to decrease its memory so that it can respond more quickly when the distribution drift occurs at a delay from when monitoring begins.


\end{abstract}

\section{Introduction}
\label{sec:introduction}

API economy \cite{apiEconomy} is driving the digital transformation of business applications across
the cloud and edge environments. In such applications, API interfaces enable meeting specific business needs, such as validating the credit of a customer. By composing the API interfaces from multiple vendors a greater business value is obtained. For example, in an online bank, putting the APIs together allows supporting the different functions of the bank, such as creating accounts, providing saving programs and loans. 

The composition of APIs in such a way increases the number of combinations exponentially.  In fact, the number of combinations is unbounded as there is little limitation on how APIs can be sequenced.  More importantly, APIs are combined in unexpected ways and it is hard for the vendors providing the APIs to anticipate and test the way the APIs will be used.  This challenge exists even if only one vendor is used, due to the exponential number of possible combinations.  In addition, teams developing for the API economy many times use continuous delivery.  In the literature survey \cite{DBLP:journals/corr/ShahinBZ17}, low test coverage is highlighted as one of the challenges of continuous delivery.

This work focuses on analyzing the way APIs are composed, and identifying new ways in which the APIs are composed that were never tested before.  Such analysis enables the system analysts to be ahead of the curve and proactive in testing anticipated way in which APIs are composed and used.  Specifically, we use a Bayesian sequential technique from \cite{lindon2020sequential} to model an observed stream of API calls by the frequency of categories (i.e., APIs), either as individual APIs or as pairwise calls between one API microservice and another.  

An alert is raised if the categorical frequency distribution of observed calls differs (i.e., has drifted) from that expected under a pre-specified baseline (typically based on a previous stable sample of calls.  This Bayesian test overcomes known challenges in sequential monitoring in that the statistical guarantee of correctness of a decision of whether the distribution has changed is adversely affected by repeated monitoring and testing of drift vs the baseline; here, the guarantee is maintained no matter how long the system is monitored.  We prove the effectiveness of this technique through simulations based on real systems, where we demonstrate both that we avoid detecting drift if the observations are from the same distribution as the baseline (control of false positive error rate) at a known rate, and also that when the distortion of the observed distribution is increased relative to the baseline, that it is correctly detected (true positive rate) at a rate increasing with the degree of distortion, represented by a distribution mixing parameter.

A further issue is that if distribution drift occurs at a delay after beginning monitoring, the drift detection ability is reduced due to the memory of the test, which takes into account the full monitoring history and not just the most recent observations.  In Section~\ref{sec:memory}, we mention several modifications, which give higher weight to recent observations, in modeling the distribution of API calls, and thus, in making our drift decision.  These modifications are left for future work.



\section{Background \label{sec:background}}
A recent work \cite{chen2021did} looked at changes in confusion matrix of APIs related to machine learning models to detect shift in API performance. They proposed an adaptive sampling approach to estimate shifts in confusion matrix, requiring 90\% less data than random sampling.
\cite{waseem2021design} is a comprehensive report on designing, monitoring, and testing of microservices based on survey results and interviews with microservices practitioners.  Monitoring and logging are integral parts of microservices. Our solution can leverage monitoring logs for API usage.
\cite{delia2021designing} develops a robust API tracer to protect against adversarial attacks on APIs.

\section{Problem setup}
\label{sec:problem}

In this work, we outline some approaches for modeling patterns of API calls and detecting drift in them.  As the problem can have many aspects (e.g., the parameter inputs to the APIs, timestamp of execution, etc.), we choose to limit ourselves to modeling the patterns of which APIs are called, that is, of modeling the frequency distribution of the API calls.  As such, the problem can be described as modeling of a random variable with a categorical distribution, with values being the API URL labels. Though we focus on the application to APIs, we note that this framework can be used for similar problems of a categorical nature.

Let $f_i$ be a given API name (categorical value), such as ``$\texttt{/home\_server/data/delete}$", and $A=\{f_1,\:f_2,\dots,\:f_{|A|}\}$ be the set of potential API names under consideration.  Let $z_t$ be an observation made at time index $t\in\{1,2,\dots\}$, with $\mathbf{z}=\{z_t\}_{t=1,\dots}$ referring to the whole stream of observations, and $\mathbf{z}_t$ denoting a subset $\{z_i\}_{i=1}^t$, up to a particular time index $t$.  We will consider two observation settings, where each $z_t$ represents either a single API or pair of APIs, as follows; this can also be extended to combinations of more than 2 APIs.  Ultimately, the modeling strategy is essentially the same, with minor modifications:

\begin{enumerate}
    \item \textbf{Single APIs}: here, each $z_t$ is a single element from the set $A$, which has $|A|$ elements.  
    \item \textbf{Pairs of APIs}: $z_t = (f_i,\:f_j)$, an ordered pair, where $f_i,f_j\in A$.  We also allow parent-less or child-less pairs, where $\emptyset$ denotes a missing value.  Thus, we also allow $z_t=(\emptyset,\:f_i)$ or $(f_i,\:\emptyset)$, where $f_i\in A$, represent parent-less or child-less calls, respectively.  Denote the set of potential calls as $\mathcal{C}=(A\bigcup\{\emptyset\})\times(A\bigcup\{\emptyset\})$; this set has $|\mathcal{C}|=(|A|+1)^2$ elements.  For simplicity, we include the pair $(\emptyset,\:\emptyset)$, which is not allowed, as an element of $\mathcal{C}$ with probability 0.
    Some other pairs $(f_i,\:f_j)\in \mathcal{C}$ may not actually be possible due to constraints of the system (e.g., some APIs may not call others), and will have probability 0 in reality.
\end{enumerate}

In this work, we assume all relevant aspects of interest of the API behaviors can be characterized by the observed sequence of calls $z_1,\:z_2,\dots$.  We do not consider, for instance, the clock time gaps between observations, or associated API input hyper-parameters, but only the relative frequencies of each category $\boldsymbol{\ell}$ observed in the sequence $\mathbf{z}$.  We consider such extensions in Section~\ref{sec:future_work}.

In our modeling of the category frequencies, we assume the observed values $z_t\in\mathbf{z}$ are mutually independent and identically distributed (iid).  That is, that any given single or set of draws do not impact the likelihood of certain values being observed at other time indices.  The identical distribution assumption means that the distribution, in terms of the likelihood of each individual category value being drawn, stays stable across $t$; this makes it more sensible to model, since a moving target (changing distribution) is difficult to model without adjustments described in Section~\ref{sec:memory}.    Our task it to detect at some point $t^*$, whether the observed stream $\{z_1,\dots,z_{t^*}\}$ appear to have drifted in distribution relative to a pre-determined baseline distribution.  The set $A$ will initially have known finite size according to the values observed in the baseline, but can be extended if new APIs are observed over time.

Let $K$ now, in general, be the number of unique potential category labels $\boldsymbol{\ell}=\{\ell_1,\dots,\ell_K\}$.  That is, in the single case each category $\ell_i$ is a single API, and in the pairs case, $\ell_i$ is an ordered pair, an element of $\mathcal{C}$; so $K=|A|$ or $(|A|+1)^2$, as appropriate.

In either case, each observed value $z_t$ is modeled as an iid draw from a multinomial distribution.  A multinomial distribution is denoted $\mathcal{M}(\boldsymbol{\theta},\:n)$ where $\boldsymbol{\theta}=\begin{bmatrix}\theta_1 & \dots & \theta_K\end{bmatrix}$ is a $K$-element weight vector where $0\leq \theta_i\leq 1,\:\forall i=1,\dots,K$ and $(\sum_{i=1}^K \theta_i)=1$; sometimes the notation $\mathbf{p}$ is used instead of $\boldsymbol{\theta}$.  The multinomial distribution defines a distribution over $K$ specified elements---say, the category labels in the vector $\boldsymbol{\ell}$---with respective indices $i=1,\dots,K$; each $\theta_i$ is the probability of the $\nth{i}$ element $\ell_i$ occurring on each trial.  $n$ is a positive integer representing a number of trials, or draws.  A draw $\mathbf{x}\sim \mathcal{M}(\boldsymbol{\theta},\:n)$ from this distribution is a $K$-length vector $\mathbf{x}$ whose elements are non-negative integers and sum to $n$.  The $\nth{i}$ element of $\mathbf{x}$ is the total number of occurrences (frequency) of the element $\ell_i$, out of $n$ trials, in this random draw.  Category $\ell_i$ has an expected frequency of $n\theta_i$, on average. 

If $n=1$, a draw $\mathbf{x}$ will have a single element---say, the $\nth{i}$---being 1, and all others being 0.  This represents a random draw of the category $\ell_i$ out of the $K$ possible, that is, what is typically referred to as a draw from a categorical distribution.  We thus model each $z_t\sim \mathcal{M}(\boldsymbol{\theta},\: n=1)$, where the probability vector $\boldsymbol{\theta}$ is identical for all draws.  

Furthermore, for $\mathbf{z}_t$, consisting of $t$ total calls, let $\overline{\mathbf{z}_t}$ denote a $K$-length vector of non-negative integers summing to $n$, where the $\nth{i}$ element is the total number of occurrences of the category $\ell_i$, out of $K$.  Thus, we can model $\overline{\mathbf{z}_t}\sim\mathcal{M}(\boldsymbol{\theta},\:n=t)$; the observed $\mathbf{z}_t$ is one possible ordered sequence of realizations of the frequency totals $\overline{\mathbf{z}_t}$.  If $t=1$, then $\overline{\mathbf{z}_1}=\begin{bmatrix}0&\dots &1 &\dots & 0\end{bmatrix}$ is a unit vector of 0s and a single element with 1, as above.

\section{Drift simulation}
\label{sec:drift_simulation}

Here, we introduce notation for our simulations, which verify the statistical correctness of technique in detecting drift under the modeling assumptions.  Our simulations of observations are made from observed samples of calls on real API systems.  Say we observe $n$ calls, either in the single or pairs case.  Let $\mathbf{H}$ be a vector or matrix whose entries are all in $[0,\:1]$ and sum to 1, and thus represent a probability distribution over API category labels, as follows:

\begin{enumerate}
    \item \textbf{Single APIs}: $\mathbf{H}=\begin{bmatrix}h_1,\dots,h_K\end{bmatrix}$, where $h_i$ is the relative frequency (observed fraction or probability) of API label $f_i\in A$ being observed, out of $n$.
    \item \textbf{Pairs of APIs}:
    $\mathbf{H}_{(|A|+1)\times (|A|+1)}=\begin{bmatrix}h_{i,j}\end{bmatrix}$, where $h_{i,j}$ is the relative frequency of API pair $(f_i,\:f_j)$ occurring, out of $n$.  We can consider $\emptyset$ as an extra  $\nth{(|A|+1)}$ element, so then $h_{|A|+1,\: |A|+1}=0$ is the probability of $(\emptyset,\:\emptyset)$ occurring, which is impossible.  $\mathbf{H}$ is typically very sparse (mostly 0s), since many theoretical potential pairs are not observed, since most APIs may in practice not call each other, or the combination may be incompatible (impossible).  Each $h_{i,j}$ corresponds to one of the $K$ categories.
\end{enumerate}

In either the single or pairwise case, let $\mathbf{H}$ to represent the baseline distribution (based on real observed behavior), and let $\mathbf{H}^*$ be another distribution from which we simulate observations $z_t\sim \mathcal{M}(\mathbf{H}^*,\:n=1)$ as a single category with probabilities given by $\mathbf{H}^*$.   
Drift happens when the distribution $\mathbf{H}^*$ of the observed $\{z_t\}$, either as single or pair calls, is determined to be significantly different from the baseline $\mathbf{H}$. 

In our simulations, we create $\mathbf{H}^*=(1-\pi_t)\mathbf{H} + \pi_t\mathbf{H}'$, for some $0\leq \pi_t \leq 1$, $t=1,2,\dots$ as a mixture distribution of $\mathbf{H}$ and some other $\mathbf{H}'$, so that we can control the amount of drift by varying the mixing constants $\pi_t$ (see similar setup in \cite{ADFRZ20}). In the present work, instead of allowing $\pi_t$ to vary as in \cite{ADFRZ20}, we set it to be a constant $0\leq c \leq 1,\:\forall t$. In reality (i.e., outside of simulations), $\mathbf{H}^*$ will be some unknown distribution.  

In general, though, $\pi_t$ can vary over time, as in the scenarios in Figure 5 of \cite{ADFRZ20}.  If $\pi_t=0$, $\mathbf{H}^*=\mathbf{H}$, so there is no drift.  In \cite{ADFRZ20}, we allowed $\mathbf{H}^*$ to shift gradually from $\mathbf{H}$ to $\mathbf{H}'$ by letting $\pi_t$ increase gradually from 0 to 1.  Furthermore, drift was allowed to begin at a delay of $t_s-1$ time points, where $t_s\geq 1$ is the first index for which $\pi_t>0$, when drift is first introduced; drift ended at some time $t_e\geq t_s$, at which $\pi_t=1$.  That is, for $t<t_s$, draws of $z_t$ would come from $\mathbf{H}$, and for $t\geq t_e$, they come from $\mathbf{H}'$.  If $t_s=1$, drift begins immediately, otherwise there is a delay while the initial draws are still from $\mathbf{H}$.  For $t_s\leq t\leq t_e$, we may increase $\pi_t$ in a linear or other fashion to introduce the drift gradually.

The ability to detect drift should depend on how significantly $\mathbf{H}$ differs from the mixture $\mathbf{H}^*$; this depends both on how $\mathbf{H}$ and $\mathbf{H}'$ differ, and on the value of $\pi_t$.  For fixed $\mathbf{H}'$, the difference from $\mathbf{H}$ increases as $\pi_t$ grows.  

\section{Sequential identification of drift of an API stream}
\label{sec:sequential}

\subsection{The importance of sequential analysis \label{ssec:sequential_analysis}}

As mentioned in Section~\ref{sec:problem}, our problem consists of observing an ordered sequence $\mathbf{z}=\{z_1,\:z_2,\dots\}$, each potential value of which is either a single or an ordered pair of APIs.  In either case, we can consider each potential value, even the pairs, as a unique category label.  On the basis of the observed API calls, we would like to decide if the sequence seems significantly anomalous relative to what would be expected if they were drawn from the baseline $\mathbf{H}$; this makes it a sequential decision problem.

As mentioned in our prior work (\cite{ADFRZ20}), performing sequential decisions is more complicated to do in an appropriate statistical way than it is to make a single non-sequential decision on a sample of values.  In short, we typically want to be able to peek at the data (e.g., the sequence $\{z_1,\dots,z_t\}$ up to time $t$, or some moving window subset of it) to make as timely a decision of drift as possible, without waiting to observe the `entire' sequence. 

Traditional (non-sequential) hypothesis tests or decision problems typically have a statistical guarantee, such as on the type-1 error (false alarm rate) that assumes a single test or independence requirements.  However, in sequential decision-making, we may want a statistical guarantee on our \textit{single decision}, if we make it, that drift has occurred; this means we cannot naively apply methods designed for single hypothesis tests and expect the statistical guarantees to hold.  Our work in \cite{ADFRZ20} gave one example of a method (CPMs, or change point models), designed to overcome this challenge for univariate data, without parametric assumptions.

\subsection{Multinomial sequential test and Bayes Factor (BF) \label{ssec:multinomial_test}}

This work will directly apply a Bayesian technique from \cite{lindon2020sequential}.  Although \cite{lindon2020sequential} deals specifically with sequentially-observed categorical data, rather than numeric, the essential background of this technique applies to any sequential data that can be parametrically modeled.  

In Bayesian analysis, typically a parametric model is proposed for a given set of observed data.  In our case, we assume the frequencies of different single or paired API calls can be modeled by a multinomial distribution (Section~\ref{sec:problem}) with a certain probability vector $\boldsymbol{\theta}=\begin{bmatrix}\theta_1 & \dots & \theta_K\end{bmatrix}$, which is the parameter of interest.  In the Bayesian framework, this parameter of interest is modeled itself by another distribution.  In this case, typically the Dirichlet distribution is used.  The Dirichlet distribution is defined by a $K$-length hyperparameter vector $\boldsymbol{\alpha}=\begin{bmatrix}\alpha_1 & \dots \alpha_K\end{bmatrix}$, where $\alpha_i > 0,\:\forall i=1,\dots,K$.  A draw from the Dirichlet distribution $\mathbf{x}\sim\mathcal{D}(\boldsymbol{\alpha})$ is a $K$-length vector $\mathbf{x}=\begin{bmatrix}x_1 & \dots & x_K\end{bmatrix}$, where each $0\leq x_i\leq 1,\: \forall i=1,\dots, K$ and $(\sum_{i=1}^K x_i)=1$, and element-wise expected value $\textrm{EV}(x_i)=\frac{\alpha_i}{\sum_{j=1}^K \alpha_j}$.  Thus, the Dirichlet distribution models a probability weight vector, such as $\boldsymbol{\theta}$ of the multinomial distribution, where each element (weight) $x_i$ (or $\theta_i$) is proportional to its respective $\alpha_i$; furthermore, the variance of each $x_i$ decreases as $\sum_i \alpha_i$ increases.

Thus, the Bayesian model for $\boldsymbol{\theta}$, the probability vector (or matrix) of categories, is that the observed category frequencies are $\overline{\mathbf{z}_t}\sim\mathcal{M}(\boldsymbol{\theta},\:n=t)$, and then  $\boldsymbol{\theta}\sim\mathcal{D}(\boldsymbol{\alpha})$.  Modeling the (assumed unknown) $\boldsymbol{\theta}$ which governs the frequencies of categories $\overline{\mathbf{z}_t}$ is done by estimating the parameter $\boldsymbol{\alpha}$ of the Dirichlet distribution on $\boldsymbol{\theta}$.  

In the Bayesian framework, the user specifies initial values for this parameter $\boldsymbol{\alpha}$ (called the prior distribution), and the parameter values are sequentially updated in light of the observed data (posterior distribution).  For instance, here we will specify the prior distribution ($\boldsymbol{\alpha}_0$, where subscript `$0$' indicates the values at time $t=0$) with respective values $\alpha_{0,i}$ being proportional to the values in the baseline $\mathbf{H}$; categories $z_t$ that are observed but not in the baseline (e.g., new APIs not in $A$, or potential pairs in $\mathcal{C}$ which had 0 frequency in $\mathbf{H}$) are given very small (close to 0 but not exactly 0) prior weights.   

At time $t$, the current posterior is $\boldsymbol{\alpha}_t$, and both the prior and posterior distributions are Dirichlet. The updating rules for the Dirichlet are that if $z_t$ corresponds to category indexed $i$, the new posterior value is the previous increased by 1 (e.g., $\alpha_{t,i}=\alpha_{t-1,i}+1$).  In this case, we say that the Dirichlet distribution is conjugate for the multinomial, because after updating when the data are multinomial-distributed, the posterior remains Dirichlet-distributed.

The posterior value $\boldsymbol{\alpha}_t$ takes into account both the prior $\boldsymbol{\alpha}_0$ and updates from the observed data $\mathbf{z}_t$.  Given any appropriate vector value for $\boldsymbol{\alpha}$ and set of observed values $\mathbf{z}$ of length $n$, we can calculate how well $\boldsymbol{\alpha}$ seems to characterize $\mathbf{z}$, by evaluating the likelihood function $L$ of the Dirichlet with both $\mathbf{z}$ (in the form of frequencies of counts $\overline{\mathbf{z}}$) and $\boldsymbol{\alpha}$; this applies to any parametric distribution, not just the multinomial.  The higher $L(\boldsymbol{\alpha},\:\overline{\mathbf{z}})$, the better the fit.

In particular, we can evaluate the relative fits of the prior $\boldsymbol{\alpha}_0$ and posterior $\boldsymbol{\alpha}_t$ at any given $t$, to determine if drift has happened; drift in this case means the posterior fits the data much better than the prior does.  This is done by forming the ratio, or odds, $L(\boldsymbol{\alpha}_t,\:\overline{\mathbf{z}}) / L(\boldsymbol{\alpha}_0,\:\overline{\mathbf{z}})$, called the posterior odds.  A user can specify prior odds, a (subjective) guess as to how much to pre-favor the prior $\boldsymbol{\alpha}_0$; for instance, prior odds of 2 (i.e., $1\colon2$) means the user wants to give twice as much weight to the prior's fit over the posterior.  Usually, the prior odds are set to 1 to be agnostic.  The Bayes Factor (BF) is the product of the posterior and prior odds, and evaluates the relative fit of the prior and posterior, while taking into account the user's confidence in the initial prior.  Since likelihood functions $L$ and odds must always be positive-valued, we have that $\textrm{BF}>0$.

If the $\textrm{BF}>1$, the posterior fits the data better than the prior; if $0<\textrm{BF}<1$, the reverse is true, and if the $\textrm{BF}\approx 1$, both fit about the same.  For a given constant $k>1$, if the $\textrm{BF}=k$ (say, $k=50$), this means the posterior fit is $k$ times better than the prior; alternatively, since we can take the reciprocal of the BF, a $\textrm{BF}=1/k$ means the prior fit is $k$ times better (or posterior is $1/k$ times as good).  The higher the BF, the more confidence we have in the posterior, relative to the prior, and thus the distribution drift appears to be more extreme.  

As mentioned, in many statistical hypothesis decision-making setups, the user wants to primarily control the false positive rate (i.e., false detection of distribution change when it hasn't really happened) to be lower than a pre-specified level $0<\alpha<1$, not to be confused with the $\boldsymbol{\alpha}$ parameters of the Dirichlet distribution.  $\alpha$ is usually set close to 0, with lower values corresponding to a more conservative decision, in that the observed deviation has to be more extreme (i.e., improbable) relative to the null hypothesis to declare drift.  In our case, a lower $\alpha$ means the posterior distribution over categories needs to be more different from the prior expected distribution than otherwise, in order to declare the distribution has drifted.  As we now see, this has a direct connection to the measured BF.  

In \cite{lindon2020sequential}, their proposed test, which we use, has the following rule: rejecting the posterior in favor of the prior, if $\textrm{BF}>1/\alpha$, has a false positive rate of $\alpha$.  That is, say, if the $\textrm{BF}>100=1/0.01$, this decision that drift happened has a false positive rate of $\alpha=0.01$.  The key aspect here is that this decision can be made in a sequential manner without sacrificing the statistical guarantee, unlike other statistical tests whose guarantee $\alpha$ no longer holds once they are applied more than once, due to the `peeking' problem.  That is, we can observe a stream $z_1,\:z_2,\:z_3,\dots$, and evaluate the BF for the posterior $\boldsymbol{\alpha}_t$ at each $z_t$, versus the prior $\boldsymbol{\alpha}_0$ (i.e., `peek').  If the BF \textit{ever} exceeds the threshold $1/\alpha$, the decision can be made with false alarm guarantee $\alpha$.

A test that is commonly used in non-sequential testing to detect differences in distribution between categorical frequency or probability vectors is the chi-squared ($\chi^2$) test (not to be confused with the chi-squared test on contingency tables which tests independence of two different categorical variables).  The authors of \cite{lindon2020sequential} contrast their sequential Bayesian test based on the BF to a non-sequential approach of at each time $t$, performing the chi-squared test to compare the prior $\boldsymbol{\alpha}_0$ and posterior $\boldsymbol{\alpha}_t$ (normalizing each vector to sum to 1, as is required for the test).  

For a fixed decision threshold $\alpha$, the non-sequential test would decide drift at the first $t$ where the chi-squared test yielded a p-value less than $\alpha$; correspondingly, the sequential test would decide drift at the first $t$ where the $\textrm{BF}>1/\alpha$.  They show that the chi-squared test results in a false positive rate that is higher than the expected $\alpha$, precisely due to the `peeking' problem and that the decision process does not adjust for the fact that the test is performed sequentially; the multinomial test, in contrast, maintains the expected $\alpha$-level control over the false positive rate.  If the simpler non-sequential test performs just as well as the more complicated multinomial test in terms of false positive rate control, all other properties being equal, there would be no need for a more complicated procedure, hence the need to demonstrate this.  In our simulated experiments, however, applying the chi-squared test seemed to actually give similar results to the multinomial test.

\section{Results \label{sec:results}}
\subsection{Simulation setup \label{ssec:sim_setup}}

As noted in Section~\ref{sec:drift_simulation}, we simulate distribution drift from $\mathbf{H}$ to another $\mathbf{H}'$ by forming a mixture distribution $\mathbf{H}^*$ of the two, with mixing parameter $0\leq \pi_t\leq 1$ at time $t$.  Setting $\pi_t=0,\:\forall t$ is thus no drift, since $\mathbf{H}^*=\mathbf{H},\:\forall t$.  Thus, when applying the multinomial test from \cite{lindon2020sequential} (Section~\ref{ssec:sequential_analysis}), at any given value $0<\alpha<1$, we should expect a false drift detection probability of $\alpha$, if we detect drift when the BF exceeds $1/\alpha$.  This can be verified by multiple repetitions of simulated sequences from $\mathbf{H}^*=\mathbf{H}$, and seeing that at most $(100\alpha)$\% of sequences ever have the BF exceed $1/\alpha$, at any $\alpha$ we choose, when the prior $\boldsymbol{\alpha}_0$ reflects the baseline $\mathbf{H}$.  Furthermore, we should see that at any $\alpha$, when we \textit{do} introduce drift (i.e., $\pi_t>0$, so $\mathbf{H}^*\ne\mathbf{H}$), we should see that \textit{more than} $(100\alpha)$\% of simulated sequences are detected to have drifted, and that this proportion should increase with increasing amounts of drift (larger difference between $\mathbf{H}^*$ and $\mathbf{H}$, which for a fixed second distribution $\mathbf{H}'$, means increasing $\pi_t$). 

In these experiments, we will illustrate only the pairwise API setting.  We begin with two frequency matrices $\mathbf{F}$ and $\mathbf{F}'$ below, which represent $n=89$ and 88 observations, respectively. These matrices represent calls on the same system, so the sets of APIs covered are the same, and frequencies are fairly similar.  The most frequent pairs are (frontend, currencyservice) and (frontend, productcatalogservice).

\begin{equation*}
\begin{aligned}
    \mathbf{F}=
&\begin{blockarray}{rrrrrrrrrrr}
\begin{block}{r(rrrrrrrrrr)}
  \emptyset & 0 &   0 &   0 &   0 &   0 &   0 &   0 &   0 &   0 &   0 & \\
  \textrm{adservice} & 0 &   0 &   0 &   0 &   0 &   0 &   0 &   0 &   0 &   0 & \\
  \textrm{cartservice} & 0 &   0 &   0 &   0 &   0 &   0 &   0 &   0 &   0 &   0 & \\
  \textrm{checkoutservice} & 0 &   0 &   0 &   0 &   2 &   0 &   0 &   1 &   0 &   0 & \\
  \textrm{currencyservice} & 0 &   0 &   0 &   0 &   0 &   0 &   0 &   0 &   0 &   0 & \\
  \textrm{frontend} & 0 &   2 &   9 &   0 &  17 &   0 &   0 &  38 &   0 &   2 & \\
  \textrm{loadgenerator} & 0 &   0 &   0 &   0 &   0 &  10 &   0 &   0 &   0 &   0 & \\
  \textrm{productcatalogservice} & 0 &   0 &   0 &   0 &   0 &   0 &   0 &   0 &   0 &   0 & \\
  \textrm{recommendationservice} & 0 &   0 &   0 &   0 &   0 &   0 &   0 &   8 &   0 &   0 & \\
  \textrm{shippingservice} & 0 &   0 &   0 &   0 &   0 &   0 &   0 &   0 &   0 &   0 & \\
 \end{block}
\end{blockarray}\\
\mathbf{F}'=
&\begin{blockarray}{rrrrrrrrrrr}
\begin{block}{r(rrrrrrrrrr)}
 \emptyset & 0 &   0 &   0 &   0 &   0 &   0 &   0 &   0 &   0 &   0 & \\
  \textrm{adservice} & 0 &   0 &   0 &   0 &   0 &   0 &   0 &   0 &   0 &   0 & \\
  \textrm{cartservice} & 0 &   0 &   0 &   0 &   0 &   0 &   0 &   0 &   0 &   0 & \\
  \textrm{checkoutservice} & 0 &   0 &   1 &   0 &   0 &   0 &   0 &   0 &   0 &   1 & \\
  \textrm{currencyservice} & 0 &   0 &   0 &   0 &   0 &   0 &   0 &   0 &   0 &   0 & \\
  \textrm{frontend} & 0 &   5 &   6 &   0 &  22 &   0 &   0 &  34 &   2 &   3 & \\
  \textrm{loadgenerator} & 0 &   0 &   0 &   0 &   0 &   7 &   0 &   0 &   0 &   0 & \\
  \textrm{productcatalogservice} & 0 &   0 &   0 &   0 &   0 &   0 &   0 &   0 &   0 &   0 & \\
  \textrm{recommendationservice} & 0 &   0 &   0 &   0 &   0 &   0 &   0 &   7 &   0 &   0 & \\
  \textrm{shippingservice} & 0 &   0 &   0 &   0 &   0 &   0 &   0 &   0 &   0 &   0 & \\
 \end{block}
\end{blockarray}
\end{aligned}
\end{equation*}

\begin{figure}
    \centering
    \includegraphics[width=0.49\linewidth]{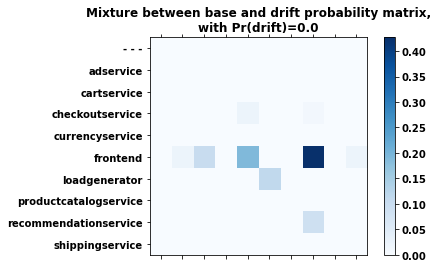}
    \includegraphics[width=0.49\linewidth]{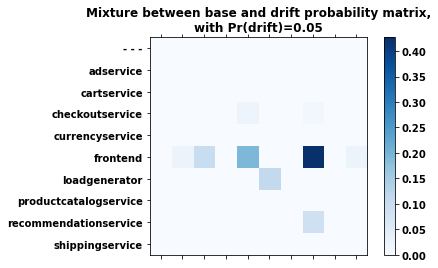}
    \includegraphics[width=0.49\linewidth]{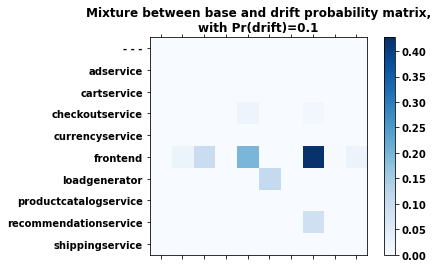}           \includegraphics[width=0.49\linewidth]{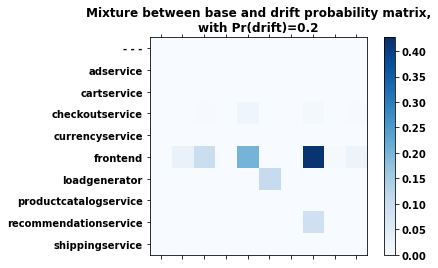}
    \includegraphics[width=0.49\linewidth]{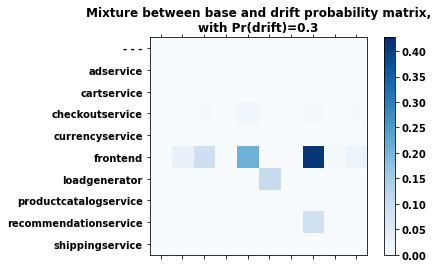}
    \includegraphics[width=0.49\linewidth]{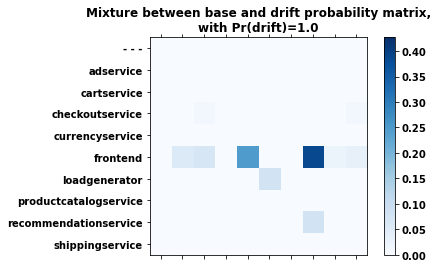}
    \caption{\label{fig:mixture_matrices} Mixture probability matrices $\mathbf{H}^*$ of $\mathbf{H}$ and $\mathbf{H}'$, with varying mixing proportions $\pi$.  Note, the mixture images seem identical, but they are not; they only appear so because the underlying distributions $\mathbf{H}$ (top left) and $\mathbf{H}'$ (bottom right) themselves are very similar.
}
\end{figure}

In our simulations, we experiment with various drift contamination proportions of $\pi=\{0.0,\: 0.05,\: 0.10,\: 0.20,\: 0.30,\: 1.0\}$ on the resulting probability matrices $\mathbf{H}$ and $\mathbf{H}'$ (each of which is $\mathbf{F}$ or $\mathbf{F}'$ divided by the sum of entries $n$).  $\pi=0$ and $1$ correspond to $\mathbf{H}$ and $\mathbf{H}'$, the baseline and alternate distributions, shown in the upper left and lower right of Figure~\ref{fig:mixture_matrices}, respectively.  Because the original frequency matrices are similar, the mixture images do not seem very different, making our drift detection more challenging.

In the sequential multinomial test, the posterior $\boldsymbol{\alpha}_t$ of the Dirichlet distribution is compared to the prior $\boldsymbol{\alpha}_0$ (Equation~\ref{eq:prior}).  A prior value $\alpha_{0,i}=0$ for the category $\ell_i$ represents a prior belief that this category is impossible.  Therefore, if this category ends up being observed (so the posterior $\alpha_{t,i}>0$), the BF would be infinite since the denominator would be 0.  Hence, we give each unobserved category (0 frequency in $\mathbf{F}$)
in the baseline a small positive prior value rather than 0.  Our matrix of prior probability values $\boldsymbol{\alpha}_0$, shown below, is calculated by rescaling the nonzero elements of $\mathbf{H}$ to sum to $n=50$ (the weight we give the prior baseline sample), then setting the zero-valued frequency elements to have a low weight of $c'=0.00006$ (the value chosen so $c'n/(\textrm{\# nonzero elements})\leq c$, for $c=0.0001$, to avoid giving them too much total weight.
 
 \begin{equation}
 \label{eq:prior}
     \boldsymbol{\alpha}_0=
 \tiny{
\begin{blockarray}{rrrrrrrrrrr}\\
\begin{block}{r(rrrrrrrrrr)}
 \emptyset & 0.0  &     0.00006 & 0.00006 & 0.00006  & 0.00006 &  0.00006 &  0.00006 &  0.00006& 
  0.00006 &  0.00006\\
  \textrm{adservice} & 0.00006&   0.00006 &  9.55056 &  1.1236 &   5.05618 &  0.00006 &  0.00006 &  1.1236& 
   0.00006 & 21.34831\\
  \textrm{cartservice} &0.00006 & 0.00006 &  0.00006  & 0.00006 &  0.00006 &  0.00006 &  0.00006 &  0.00006 & 
   0.00006 &  0.00006\\
 \textrm{checkoutservice} & 0.00006 &  0.00006 &  0.00006 &   0.00006 &  0.00006 &  0.00006 &  0.00006 &  0.00006 & 
   0.00006 &  0.00006\\
 \textrm{currencyservice} & 0.00006 &  0.00006 &  0.00006 &  0.00006 &  0.00006 &  0.00006 &  0.00006 &   0.00006& 
   0.00006 &  0.00006\\
 \textrm{frontend} & 0.00006 &  5.61798 &  0.00006 &  0.00006 &  0.00006 &  0.00006 &  0.00006 &  0.00006 & 
   0.00006 &  0.00006\\
\textrm{loadgenerator} &  0.00006 &  0.00006 &  1.1236  &  0.00006 &  0.00006 &  0.00006 &  0.00006 &   0.00006 & 
   0.00006 &  0.5618 \\
\textrm{productcatalogservice} &  0.00006  & 0.00006 &  0.00006 &  0.00006 &  0.00006 &  0.00006 &  0.00006 &  0.00006 & 
   0.00006 &  0.00006\\
\textrm{recommendationservice} &  0.00006 &  0.00006 &   0.00006 &  0.00006 &  0.00006 &  0.00006 &  0.00006 &  0.00006 & 
   0.00006 &  4.49438\\
\textrm{shippingservice} &  0.00006 &  0.00006 &  0.00006 &  0.00006 &  0.00006 &  0.00006 &  0.00006 &  0.00006 & 
   0.00006 &  0.00006\\\\
 \end{block}
\end{blockarray}
}
\end{equation}

\subsection{Simulation with no drift\label{ssec:sim_nodrift}}

\begin{figure}
    \centering
    \includegraphics[width=0.85\linewidth]{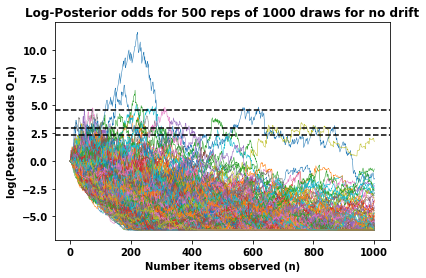}
    \caption{\label{fig:nodrift}Log BF for $r=500$ repetitions of $n=1,000$ random draws from $\mathbf{H}^*=\mathbf{H}$ (no drift).
}
\end{figure}

To verify that the statistical guarantees are satisfied, we generate $r=500$ repetitions of $n=1,000$ draws for each simulation. In the first, we set $\pi_t=0,\:\forall t$.  We will test the detection with confidence thresholds of $\alpha=\{0.10,\:0.05, 0.01\}$.  As noted in Section~\ref{ssec:sequential_analysis}, the drift detection thresholds for the BF are $1/\alpha$.  In Figure~\ref{fig:nodrift}, we show the natural logs of BF for each repetition (each in a different color), as well as the log thresholds as horizontal black dashed lines at $2.3026,\: 2.9957,\: 4.6052$.  The first row of Table~\ref{tab:detection_rates}, shown in Section~\ref{ssec:sim_drift}, shows the false alarm rate is controlled, since at each $\alpha$, the proportion of simulations in which drift is (falsely) detected is approximately $\alpha$.

\subsection{Simulation with drift \label{ssec:sim_drift}}

Next, we simulate the same number of repetitions and time steps for each mixing proportion $\pi=\{0.05,\:0.10,\:0.20,\:0.30\}$ as in Section~\ref{ssec:sim_setup}.  Again, the log BF and thresholds are plotted.  Since drift is detected for a repetition if the BF ever passes the respective threshold, we see clearly that as $\pi$ increases, the bulk of the lines pass the thresholds at each $\alpha$; at each $\alpha$, this proportion should be greater than $\alpha$, and increase with $\pi$, as shown in Table~\ref{tab:detection_rates}.

\begin{figure}
    \centering
    \includegraphics[width=0.49\linewidth]{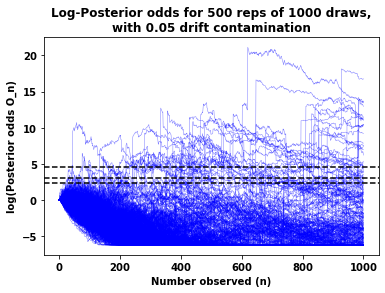}
    \includegraphics[width=0.49\linewidth]{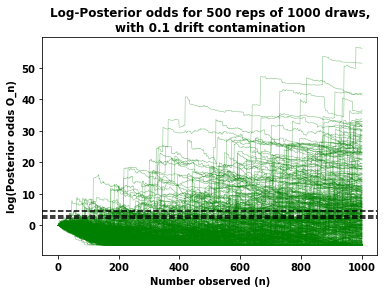}
    \includegraphics[width=0.49\linewidth]{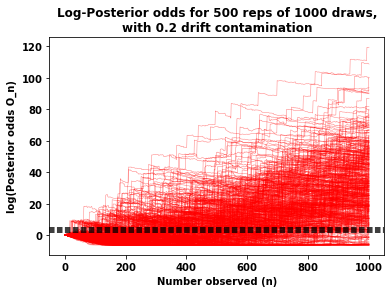}
    \includegraphics[width=0.49\linewidth]{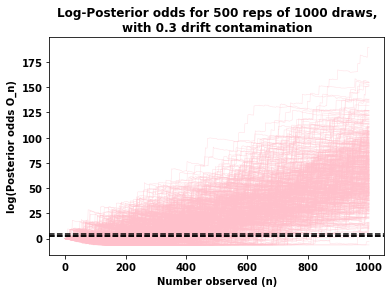}
    \caption{\label{fig:drift_sim}Log BF for $r=500$ repetitions of $n=1,000$ random draws from $\mathbf{H}^*=(1-\pi)\mathbf{H} + \pi\mathbf{H}'$ (no drift), for various values of $\pi>0$.  The detection success should increase with $\pi$.  Horizontal dashed lines show detection thresholds of the BF at values $\alpha=\{0.1,\:0.05,\:0.01\}$, from bottom to top.  
}
\end{figure} 

\begin{table}[ht]
\renewcommand{\arraystretch}{1.5}
\centering
\begin{tabular}{r|c |c| c |} 
\hline
\multicolumn{1}{c|}{Drift proportion $\pi$} & \multicolumn{3}{|c|}{Threshold $\alpha$} \\
\cline{2-4}
\multicolumn{1}{c|}{} & 0.10 & 0.05 & 0.01\\
\hline
0.0 (no drift) & 0.096 & 0.044 & 0.014\\
0.05 & 0.19 & 0.15 &  0.108\\
0.10 & 0.554 & 0.508 & 0.438\\
0.20 & 0.946 & 0.936 & 0.922\\
0.30 & 0.998 & 0.998 & 0.996\\
\hline
\end{tabular}
\caption{\label{tab:detection_rates}
Proportion of simulation repetitions where drift is detected.  When $\pi>0$, this should be the correct decision.  When $\pi=0$, the proportions should be approximately $\alpha$ (the desired false positive rate), showing that the false positive rate is controlled. 
}
\end{table}

\subsection{\label{ssec:results_analysis} Contribution of categories to drift detection}

The analyses described in the following will try to determine and which categories (e.g., pairs of APIs) are responsible for the difference between the observed and baseline distributions, and to measure either the effect on the drift decision directly, or to measure the degree of divergence from the expected frequency, either more or less than expected.  One application of drift detection to APIs is to testing of systems. For instance, a test can be run to simulate various scenarios of sets of APIs and their interactions.  Through drift testing, we can see if the test has generated a different distribution pattern of API calls (in terms of relative frequencies) than the baseline; this can help assess if the test has generated the expected results.  If a test is known to, for instance, tend to increase some API pair frequency, we can see if the measured divergence accords with this anticipated change.

For an observation $z_t$, let the operator $\nu(t)\in\{1,\dots,K\}$ denote the index of the category labels $\{\ell_1,\dots,\ell_K\}$ observed at time $t$ by observation $z_t$.  So if $z_t=\ell_i$, $\nu(t)=i$, so $z_t=\ell_{\nu(t)}$.

In \cite{lindon2020sequential} (Appendix A, equation 25), the authors show that the BF at $t$ can be calculated recursively based on the BF at $t-1$, taking into account the new observation.  In their notation, $x_{t,i}=I(z_t=\ell_i)$, where $I$ is the indicator function; that is, $x_{t,\nu(t)}=1$, while $x_{t,i}=0$ for categories $\ell_i$ not observed.  Note that the posterior $\boldsymbol{\alpha}$ is updated only for the category observed, so if $i=\nu(t)$, $\alpha_{t,i}=\alpha_{t-1,i}+1$, otherwise $\alpha_{t,i}=\alpha_{t-1,i}$.
Simplifying the multiplier of the BF at $t-1$ (in their notation, $O_{n-1}(\boldsymbol{\theta}_0$)) using the facts above, it can be written as 

\[
\psi_t=\left(\frac{\Gamma(\sum_{i=1}^K\alpha_{t-1,i})}{\Gamma((\sum_{j=1}^K\alpha_{t-1,j}) + 1)}\right)\left(\frac{\Gamma(\alpha_{t-1,\nu(t)}+1)\prod_{i\ne \nu(t)}\Gamma(\alpha_{t-1,i})}{\Gamma(\alpha_{t-1,\nu(t)})\prod_{j\ne \nu(t)}\Gamma(\alpha_{t-1,j})}\right)\left(\frac{1}{\theta_{0,\nu(t)}}\right)
\]

where $\Gamma$ denotes the gamma function, and $\theta_{0,i}$ is the null hypothesis value of $\textrm{Pr}(\ell_i)=\frac{\alpha_{0,i}}{\sum_j\alpha_{0,j}}$.  By cancellation and using the fact that $\Gamma(a+1)=a\Gamma(a)$, we further have

 \[
 \psi_t=\frac{\alpha_{t-1,\nu(t)}}{\theta_{0,\nu(t)}\sum_{i=1}^K\alpha_{t-1,i}}=\frac{\alpha_{t-1,\nu(t)}/(\sum_{i=1}^K\alpha_{t-1,i})}{\theta_{0,\nu(t)}}=\frac{\textrm{posterior probability of $\ell_{\nu(t)}$}}{\textrm{prior probability of $\ell_{\nu(t)}$}}
 \]
 
Thus the BF at time $t$ is the BF at $t-1$ multiplied by $\psi_t$.  Because of this recursive multiplication, if we define $\phi_0$ as the prior odds, we can say that the BF at time $t$ is  $\prod_{j=0}^t\psi_j$.  The denominator of $\psi_t$ is the value of $\alpha_{t-1,\nu(t)}$ if the null prior hypothesized probability $\theta_{0,\nu(t)}$ for observed category $z_t=\ell_{\nu(t)}$ was correct; the denominator is the actual posterior value.  If $\psi_t>1$, it means the posterior has given more likelihood to $\ell_{\nu(t)}$ than the null (i.e., prior), which will increase the previous time's BF; the reverse is true if $\psi_t<1$.  

Since $\ln{(a)}=-\ln{(1/a)},\:a>0$, we can consider $\ln{(\psi_t)}$ as the additive effect of $z_t$ in contributing to the BF.  Only cases where the BF is large ($>1/\alpha$), that is, when we tend to see $\psi_t>1$, contribute to the drift detection decision.  This happens when we see categories $\ell_i$ happen \textit{more often} than expected, but this also requires other categories to happen \text{less often} than expected.  Thus, if drift is detected and we stop observing at that point (i.e., the last BF value we have is the one passing the threshold), we can consider an overall measure of the influence of category $\ell_i$ in the decision to detect drift as
 
 \[\Delta(i)=\sum_{t\colon \nu(t)=i}\ln{(\psi_t)}=\sum_{t\colon z_t=\ell_i}\ln{(\psi_t)}
 \]
 
Since, by definition, categories $\ell_i$ which happen more often will have $\Delta(i)$ summed over more indices $t$, the overly-frequent categories will tend to have high $\Delta(i)$.  However, it would be possible for a category to, say receive a high $\Delta(i)$ for appearing still frequently, but less frequently than expected.  Categories $\ell_i$ which happen around as frequently as expected under the prior, whether frequent or not, will tend to have $\psi_t\approx 1$, which makes their $\Delta(i)$ tend to be close to 0.  Recall that the order of observations matter, so temporal instability in the probability of a category $\ell_i$ occurring---whether by randomness, though we assume the probabilities are iid, or representing an actual shift---can affect the detection of drift, even if on average, the category was observed at the prior expected rate.  

One weakness of the $\Delta$ metric above is that categories $\ell_i$ that are unobserved, even though they may have received high prior weight, are not counted. Therefore, we also below consider a log ratio metric $\rho(i)$, as follows. Let $\mathbf{F}$ and $\mathbf{F}'$ be the prior (expected) and observed (posterior minus prior) and frequency matrices, respectively, with entries summing to $n_0$ and $n_1$; zero-valued entries in $\mathbf{F}'$ are replaced with the same minimum constant $c$ from the prior (see Section~\ref{ssec:sim_setup}). Let $F_i$ and $F_i'$ be frequency values corresponding to a given category $\ell_i$, and define a metric 

\[\rho(i)=\ln{\left(\frac{\textrm{max}(F_i',\: 0.5)/n_1}{\textrm{max}(F_i,\: 0.5)/n_0}\right)}
\]

Since unseen categories in the prior or observations have $F_i$ or $F_i'<1$, ratios $\frac{F_i'}{F_i}$ can be unstable if one of them is unseen.  In the logarithm, we thus set them to have a minimum value of 0.5 so that, say, a category that was unseen in one but seen only once in the other would have a score of $\rho(i)=\pm\ln(2)\approx\pm0.693$, rather than much higher, to avoid giving too anomalous scores to these low-frequency categories. Categories $\ell_i$ where $\rho(i)>0$ are observed more often than expected under the prior; the opposite is true for negative values.  Thus, unlike under the $\Delta$ metric, categories $\ell_i$ that are not observed in one side ($F_i$ or $F_i'\approx 0$) can still have high (negative) anomalousness if their expected frequency $F_i$ is high enough.  The $\rho$ metric measures anomalousness in the final frequencies without regard to the order, unlike $\Delta$, which sums the $\psi_t$ anomalous scores according to the sequence APIs were observed.

\begin{figure}
    \centering
    \includegraphics[width=0.49\linewidth]{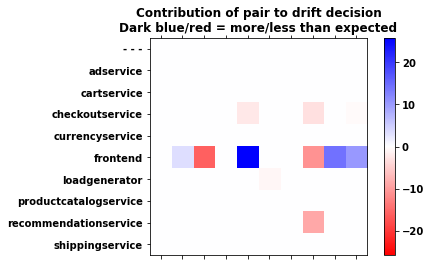}
    \includegraphics[width=0.49\linewidth]{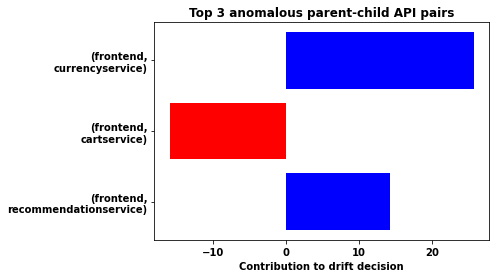}
    \caption{\label{fig:contrib_all}Contribution $\Delta(i)$ for various API pairs $\ell_i\in\mathcal{C}$.  
}
\end{figure}

To demonstrate, we now run a single simulation with $\pi=0.20$ and decision threshold $\alpha=0.01$.  This particular run detected drift after $t=559$ observations. The resulting contribution scores $\{\Delta(i)\}_{i=1}^K$ for API pairs are shown in the left plot of Figure~\ref{fig:contrib_all}.  Any grid square that is visibly shaded is contributing to the decision; darker hues mean higher contribution, and blue/red color indicates positive/negative value of the contribution $\Delta(i)$.  The right plot shows the top 3 pairs, ordered by $|\Delta(i)|$, which can be viewed as an anomaly score, regardless of the sign of $\Delta(i)$.  The third most anomalous category, (frontend, recommendationservice), was observed only 2 times, but was judged to be highly anomalous because it received the minimum weight under the prior, since it was not observed under the baseline.  Its prior probability is $\theta_{0,i}$, the prior value $\alpha_{0,i}\in\boldsymbol{\alpha}_0$ when divided by the sum of elements in $\boldsymbol{\alpha}_0$; under the prior, out of $t=559$ draws, it would be expected to be observed with essentially 0 frequency ($0.00062\approx 559\theta_{0,i}$).  The observed and expected frequencies for these pairs are shown in Table~\ref{tab:pair_topk_freqs} 

\begin{table}[ht]
\centering
\begin{tabular}{l|rr}
  \hline
Category $\ell_i$ & Observed count ($F_i'$) & Expected count ($F_i$) \\
\hline
\textrm{(frontend, currencyservice} & 128 & 106.77528 \\
\textrm{(frontend, cartservice)} & 48 & 56.52809\\
\textrm{(frontend, recommendationservice)} & 3 & 0.00062\\
\hline
\end{tabular}
\caption{\label{tab:pair_topk_freqs} Observed and expected counts for top 3 anomalous API pairs, by $\Delta$ metric.}
\end{table}

\begin{figure}
    \centering
    \includegraphics[width=0.49\linewidth]{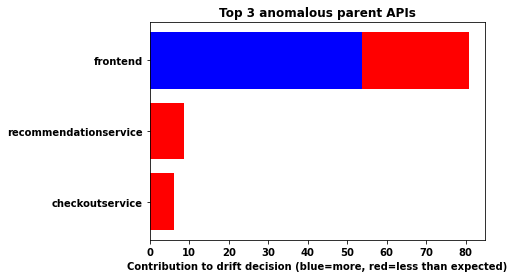}
    \includegraphics[width=0.49\linewidth]{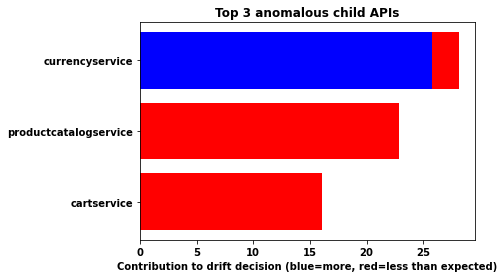}
    \caption{\label{fig:contrib_parent_child}Contribution $\Delta$ for various parent/child APIs for API pairs in $\mathcal{C}$.  
}
\end{figure}

Similarly, we can rank individual parent ($a$) and child ($b$) APIs, where $a,b\in\{\emptyset\}\bigcup A$, in terms of anomalousness, by summing 

\[\displaystyle{\sum_{i\colon\:a\textrm{ is the parent in }\ell_i}|\Delta(i)|}\quad\textrm{ or }\displaystyle{\sum_{i\colon\:b\textrm{ is the child in }\ell_i}|\Delta(i)|}
\]
respectively.  These are shown in Figure~\ref{fig:contrib_parent_child}; the same calculation can be done for $\rho$ by substituting it for $\Delta$ in the equation.  The frontend API was the most anomalous parent, for instance, because many API pairs having it as a parent (Figure~\ref{fig:contrib_all}, left plot, `frontend' row) were anomalous; in addition, its anomalousness is due both to child APIs that were called either more (blue) or less (red) often than expected, as shown the the relative lengths of blue/red bars.

\begin{figure}
    \centering
    \includegraphics[width=0.49\linewidth]{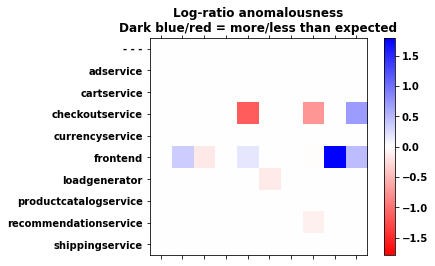}
    \includegraphics[width=0.49\linewidth]{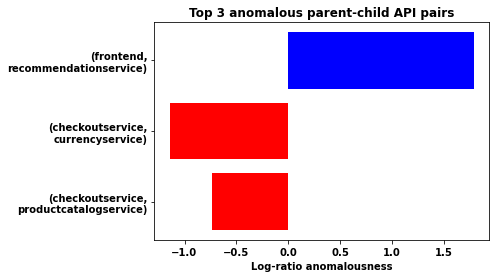}
    \caption{\label{fig:rho_all}Anomalousness $\rho(i)$ for various API pairs $\ell_i\in\mathcal{C}$.  
}
\end{figure}

\begin{figure}
    \centering
    \includegraphics[width=0.49\linewidth]{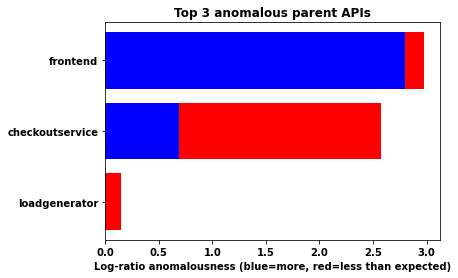}
    \includegraphics[width=0.49\linewidth]{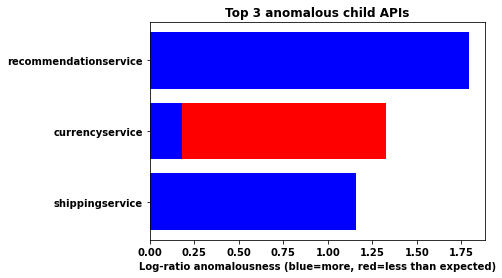}
    \caption{\label{fig:rho_parent_child}Anomalousness $\rho$ for various parent/child APIs for API pairs in $\mathcal{C}$.  
}
\end{figure}

Noting the drawback of the $\Delta$ metric above, we show the results of the $\rho$ metric on the same simulation, which can also measure anomalousness on potential categories $\ell_i$ which were not observed, and hence their absence is anomalous if they were expected.  These pairwise results are shown in Figure~\ref{fig:rho_all}.  In the left plot, we see that one pair, (frontend, recommendationservice), turns out similarly anomalous to the results in the $\Delta$ metric (Figure~\ref{fig:contrib_all}), in that it was more frequent than expected.  The observed and expected values for the top 3 pairs are shown in Table~\ref{tab:pair_topk_freqs_rho}.  The $\rho$ metric, since it is calculated on ratios, seems to emphasize differences in the relatively infrequent pairs rather than in the more frequent ones in the $\Delta$ metric.    Figure~\ref{fig:rho_parent_child} shows similar results summed across parent and child APIs.

\begin{table}[ht]
\centering
\begin{tabular}{l|rr}
  \hline
Category $\ell_i$ & Observed count ($F_i'$) & Expected count ($F_i$) \\
\hline
\textrm{(frontend, recommendationservice)} & 3 & 0.00062 \\
\textrm{(checkoutservice, currencyservice)} & 4 & 12.56180\\
\textrm{(checkoutservice, productcatalogservice)} & 3 & 6.28090\\
\hline
\end{tabular}
\caption{\label{tab:pair_topk_freqs_rho} Observed and expected counts for top 3 anomalous API pairs, by $\rho$ metric.}
\end{table}

Though the $\Delta$ metric has some drawbacks as compared to the $\rho$ metric, we note in Section~\ref{sec:future_work} that we may investigate modeling the length of the clock time gaps between pairs of APIs; for instance, the time gap to the next call may depend on the pair just called.  In this case, with a Bayesian model, this information can be incorporated into the likelihood function, and thus in to the BF.  The $\Delta$ metric would thus be able to reflect this, but the $\rho$ metric measures only the deviation between frequencies.
A naive approach would be to weight the occurrence of each category $\ell_{\nu(t)}$ by, say, the average of time gaps between it and the preceding and following calls $z_{t-1}$ and $z_{t+1}$, which would allow the use of frequency-based metrics like $\rho$.  However, particularly if API calls are irregularly spaced in time, it is likely incorrect to simply give higher weight to a call because it is more isolated in time.  Therefore, if the model considers aspects other than the relative frequency distribution of APIs (or associated hyper-parameters), such as the time gaps, the $\Delta$ metric could likely be adapted, since it depends on the likelihood functions, but the $\rho$ metric, since it depends on frequencies, may no longer be appropriate.
\section{Improving detection time by weighting past observations \label{sec:memory}}

By their nature, Bayesian posteriors should become more certain over time.  For instance, in the test above, if the observed data is very similar to the prior (e.g., if $\pi_t\approx 0$), the data will reinforce the prior, and the variance of the posterior will decrease.  This is true for the Dirichlet distribution since, as noted, the larger the sum of the $\boldsymbol{\alpha}$, the lower the variance, assuming the mean values of $\alpha_i$'s stay the same.  This means that if a long period of non-drift occurs (i.e., if drift begins at $t_s>>1$), it will be more difficult to detect drift if it then occurs, since the posterior has become `entrenched' into becoming  more and more `tight' around the baseline.  `More difficult' in this case means that to cause the BF to exceed the threshold $1/\alpha$, we need to observe more instances of drift to detect it than if the drift had started immediately ($t_s$ closer to 1), due to the increased certainty (tightness) of the posterior.  Since we want to be flexible about detecting drift whenever it occurs, this suggests we should make the posterior updates reflect the most recent data more than old data.

There are several approaches to this.  One is to add a forgetting constant $0<w<1$ when updating.  This approach, mentioned in \cite{MZ19} (page 6, under equation 4), can be applied.  In their example, a binary-valued variable (with values coded as `success' or `failure') is modeled as Bernoulli-distributed, where the relevant parameter $\rho\in[0,\:1]$ is the `success' probability on each draw.  $\rho$ is modeled as having a beta-distributed prior (which has the same domain $[0,\:1]$, as required).  The beta distribution is updated by incrementing one of its parameters each time by 1, depending on whether the draw ($r_j$) is a success or failure; this is the same as the Dirichlet posterior updates in our case (see Section~\ref{ssec:results_analysis}), since the beta distribution is the same as the Dirichlet when the number of categories $K=2$ (i.e., success or failure).  

In \cite{MZ19}, the past observations (past total counts of success or failure) are given a weight of $w\in(0,\:1]$ when updating the posterior of $\rho$. If $w=1$, this is the standard posterior update with no forgetting, in that the expected value of $\rho$ is the fraction of successes out of the total number of draws.  If $w < 1$, the procedure has forgetting, in that the past observations are given less weight (in proportion to $w$) than the newest observations in the update.  The efficient memory is $\frac{1}{1-w}$, so if $w=1$ this is infinite memory, since all past observations receive the same weight in the update equation.  In experiments we have performed, a constant $w<1$ did not appear to significantly reduce delay of drift detection within a reasonable time window, when the start $t_s$ of drift was large, where the results are essentially the same as with no forgetting (the standard $w=1$).

A similar, but more sophisticated, approach is illustrated in \cite{M18}, which similarly uses a forgetting factor $w$ but learns it dynamically rather than having it be fixed as in \cite{MZ19}.  Here, a model called a Hierarchical Adaptive Forgetting Variational Filter (HAFVF) has $w$ modeled by a beta distribution (which has domain $[0,\:1]$, like $w$).  At each point, a weighted sum of the initial prior and the current raw posterior (receiving weights $1-w$ and $w$, respectively) is formed to represent the current best estimate of the parameter of interest (in our case, the vector parameter $\boldsymbol{\alpha}$).  $w$ is dynamic; when the newest observed data appears to be significantly different than the prior (i.e., drift), $w$ increases to shift more weight to the raw posterior in the weighted update.  In simulations with repeated back-and-forth drift/no drift, the authors show the the weighted posterior adjusts quickly to reflect the distribution of the most recent observations (e.g., the drift distribution $\mathbf{H}'$), regardless of the time index $t$.  That is, the ability to adapt is not affected by the delay $t_s$, as it is in the constant forgetting factor.


\section{Future work \label{sec:future_work}}

In future work, we will attempt several extensions of the work shown here.  
First, we intend to adapt the HAFVF (Section~\ref{sec:memory}) for the multinomial-Dirichlet setup so drift that does not begin immediately can be detected with shorter delay.  Second, we may consider modeling API calls by a hierarchical model, for instance not just the API names but also specifying parametric models for API hyper-parameters---particularly if they can be discretized to multinomial to avoid assuming a particular continuous parametric distribution---or modeling the clock time gaps between calls.  A likelihood function can be calculated by assuming, say, independence between hyper-parameters.  However, this would make the model more complex; it may be particularly difficult to adapt the HAFVF calculations to this scenario.  Also, we intend to investigate further the particulars of the $\Delta$ and $\rho$ anomalousness metrics outlined in Section~\ref{sec:results}.

Furthermore, since in practice, APIs $\in A$ need not necessarily be combined only in pairs, but potentially in ordered triplets or more, the method can be extended.  In simulations where we specify a prior, we can consider all potential permutations, or some constrained subset of them, of elements in $A$, up to some maximum, give each of them a unique label in $\boldsymbol{\ell}$, and assign unobserved combinations a small positive prior constant value.  This prior value may be need to be adjusted relative to the length of the combination or the relative likelihoods of seeing combinations of a given length.  However, the notion of an API being a parent or child, as in the anomaly scoring in Section~\ref{ssec:results_analysis}, will need to be generalized from the pairwise formulation, where there are only two positions.  Perhaps, for instance, this summation can be done for an API $a\in A$ regardless of its position in the API combination sequence, or perhaps with some kind of importance weighting. 
\section{Conclusion \label{sec:conclusion}}

In this work we have illustrated an application of an existing method for sequential detection of changes in the distribution of observed categories, to the problem of detecting a change in the patterns of API calls.  The method provides a statistical guarantee on the confidence of its decision that distribution change (drift) has resulted.  We also presented several metrics to explain to the user which APIs were most anomalous, in terms of being observed much more or less frequently than expected, relative to a baseline.

\itemsep=0pt
{\footnotesize
\printbibliography
}
\end{document}